\documentclass[reprint,superscriptaddress,amsmath,amssymb,prb]{revtex4-2}

\usepackage{graphicx} 
\usepackage{dcolumn}  
\usepackage{bm}       
\usepackage{xcolor}   
\usepackage{epstopdf}

\begin{document}

\preprint{APS/123-QED}

\title{Coupling between Antiferromagnetic and Spin Glass Orders in the Quasi-One-Dimensional Iron Telluride TaFe$_{1+x}$Te$_3$ ($x$=0.25) }

\author{Y. Liu}
\affiliation{Key Laboratory of Quantum Precision Measurement of Zhejiang Province, Department of Applied Physics, Zhejiang University of Technology, Hangzhou 310023, China}

\author{J. J. Bao}
\affiliation{Key Laboratory of Quantum Precision Measurement of Zhejiang Province, Department of Applied Physics, Zhejiang University of Technology, Hangzhou 310023, China}

\author{C. Q. Xu}
\affiliation{School of Physics and Key Laboratory of MEMS of the Ministry of Education, Southeast University, Nanjing 211189, China}
\affiliation{Department of Physics and Astronomy, Michigan State University, East Lansing, Michigan 48824-2320, USA}

\author{W. H. Jiao}
\affiliation{Ningbo Institute of Materials Technology and Engineering, Chinese Academy of Sciences, Ningbo 315201, China}

\author{H. Zhang}
\affiliation{Department of Physics and Astronomy, Michigan State University, East Lansing, Michigan 48824-2320, USA}

\author{L. C. Xu}
\affiliation{Wuhan National High Magnetic Field Center, School of Physics, Huazhong University of Science and Technology, Wuhan, 430074, China}

\author{Zengwei Zhu}
\affiliation{Wuhan National High Magnetic Field Center, School of Physics, Huazhong University of Science and Technology, Wuhan, 430074, China}

\author{H. Y. Yang}
\affiliation{Anhui Province Key Laboratory of Condensed Matter Physics at Extreme Conditions, High Magnetic Field Laboratory, Chinese Academy of Sciences, Hefei
10 230031, China}

\author{Yonghui Zhou}
\affiliation{Anhui Province Key Laboratory of Condensed Matter Physics at Extreme Conditions, High Magnetic Field Laboratory, Chinese Academy of Sciences, Hefei
10 230031, China}

\author{Z. Ren}
\affiliation{Westlake University, School of Science, Hangzhou 310064, China}

\author{P. K. Biswas}
\affiliation{ISIS Pulsed Neutron and Muon Source, STFC Rutherford Appleton Laboratory, Harwell Campus, Didcot, Oxfordshire OX11 0QX, United Kingdom}

\author{S. K. Ghosh}
\affiliation{School of Physical Sciences, University of Kent, Canterbury CT2 7NH, United Kingdom}

\author{Zhaorong Yang}
\affiliation{Anhui Province Key Laboratory of Condensed Matter Physics at Extreme Conditions, High Magnetic Field Laboratory, Chinese Academy of Sciences, Hefei
10 230031, China}

\author{X. Ke}
\email{kexiangl@msu.edu}
\affiliation{Department of Physics and Astronomy, Michigan State University, East Lansing, Michigan 48824-2320, USA}

\author{G. H. Cao}
\email{ghcao@zju.edu.cn}
\affiliation{Department of Physics, Zhejiang University, Hangzhou 310027, China}

\author{Xiaofeng Xu}
\email{xuxiaofeng@zjut.edu.cn}
\affiliation{Key Laboratory of Quantum Precision Measurement of Zhejiang Province, Department of Applied Physics, Zhejiang University of Technology, Hangzhou 310023, China}

\date{\today}

\begin{abstract}
Understanding the interplay among different magnetic exchange interactions and its physical consequences, especially in the presence of itinerant electrons and disorders, remains one of the central themes in condensed matter physics. In this vein, the coupling between antiferromagnetic and spin glass orders may lead to large exchange bias, a property of potential broad technological applications. In this article, we report the coexistence of antiferromagnetic order and spin glass behaviors in a quasi-one-dimensional iron telluride TaFe$_{1+x}$Te$_3$ ($x$=0.25). Its antiferromagnetism is believed to arise from the antiferromagnetic interchain coupling between the ferromagnetically aligned FeTe chains along the $b$-axis, while the spin glassy state stems from the disordered Fe interstitials. This dichotomic role of chain and interstitial sublattices is responsible for the large exchange bias observed at low temperatures, with the interstitial Fe acting as the uncompensated moment and its neighboring Fe chain providing the source for its pinning. This iron-based telluride may thereby represent a new paradigm to study the large family of transition metal chalcogenides whose magnetic order or even the dimensionality can be tuned to a large extent, forming a fertile playground to manipulate or switch the spin degrees of freedom thereof.
\end{abstract}

\maketitle

\section{Introduction}

A spin glass is characterized by a random, yet cooperative, freezing of the correlated spins at the well defined temperature $T_f$ below which a highly metastable frozen state forms, without the typical long range magnetic order but with only the short-range correlations\cite{SBlundell}. Usually, this magnetically frustrated state originates from the disorder, or the so-called randomness (site-randomness or bond-randomness), and the competing exchange interactions. The competing exchange interactions, in turn, are generically observed in systems with randomness or magnetic anisotropy, where both ferromagnetic (FM) or antiferromagnetic (AFM) couplings may occur due to, e.g., RKKY interaction whose sign depends on the distance between the spins, or the Dzyaloshinsky-Moriya interaction, or other anisotropy in magnetic exchange. This rivaling interaction leads to high frustration and thereby multidegenerate ground states. Below $T_f$ in a spin glass, the ensemble of spins undergo a cooperative phase transition and freeze into a metastable glassy state.

Experimentally, one characteristic of this metastable state is the bifurcation in the zero-field cooled (ZFC) and field-cooled (FC) magnetization curves below $T_f$, as seen in many spin glasses such as IrMnGa\cite{Felser-IrMnGa}, Ni-Mn-X (X=Sb, Sn and In) alloys\cite{Chatterjee,APL-WuGH}, etc. Another typical signature of spin glass behavior is the peak structure in the vicinity of $T_f$ of the ac susceptibility (both its real part and imaginary part), which shifts as the frequency of the alternating magnetic field varies. Both the irreversibility between ZFC and FC magnetizations and the shifted ac susceptibility reflect the metastability nature of its underlying spin dynamics.

Exchange bias (EB), on the other hand, is also an outcome of such competing interactions. Conventional exchange bias (CEB) manifests itself as a lateral shift in the hysteresis loop, resulting in the asymmetric coercive fields, when the sample is cooled under an external magnetic field\cite{Analytis-NP,Analytis-NC}. In recent years, large EB after zero-field-cooling from the paramagnetic state, dubbed ZFC EB, was reported in a handful of materials such as Ni-Mn-In bulk alloys\cite{BMWang-PRL}, Mn$_2$PtGa\cite{Felser-PRL,Felser-NM} and La$_{1.5}$Sr$_{0.5}$CoMnO$_6$\cite{Murthy-APL,Boldrin-APL}, where the ZFC EB effect is rooted in the FM unidirectional anisotropy formed at the interface between different magnetic phases during the initial magnetization process. In this article, we shall address the CEB effect in TaFe$_{1.25}$Te$_3$ samples. The CEB field was originally proposed to arise from the exchange anisotropy across the disordered interface between FM and AFM phases where the AFM serves as the pinning layer and the exchange interaction at the FM-AFM interface act to pin the uncompensated moments in the FM layer (pinned layer)\cite{Meiklejohn}. It has long been known that the EB phenomena can also be induced by combining FM with other phases such as a ferrimagnet or a spin glass\cite{Meiklejohn,Cain,Ali-NM,Felser-NM}, the latter now taking the role of AFM layer in the FM-AFM interface as the pinning source. However, the detailed microscopic mechanism by which this pinning drives the observed EB remains as yet elusive. Very recently, a giant EB field was observed in the magnetically intercalated transition metal dichalcogenide Fe$_{1/3+\delta}$NbS$_2$ owing to the coexisting antiferromagnetic and spin glass orders, where the spin glass provides the source of uncompensated moment which is pinned by its antiferromagnetic order\cite{Analytis-NP}.

In this article, we report the coexistence of antiferromagentic and spin glass orders in the quasi-one-dimensional (Q1D) iron-based telluride TaFe$_{1+x}$Te$_3$ ($x$=0.25). The crystal structure consists of the one-dimensional FeTe chains extended along the crystallographic $b$-axis, with Fe spins aligned ferromagnetically along the chains that are coupled antiferromagnetically in perpendicular directions\cite{Ke}. The excess amounts of Fe $x$ fill the interstitials that are responsible for the spin glass behaviors at lower temperatures. The exchange coupling between antiferromagnetic and spin glass orders gives rise to the large conventional exchange bias. Distinctly different from Fe$_{1/3+\delta}$NbS$_2$ where the dilute ($\delta$$<$0) or excess ($\delta$$>$0) irons are embedded in the triangular lattices of iron\cite{Analytis-NP}, the spins responsible for AFM and spin glass landscapes in TaFe$_{1+x}$Te$_3$ come from different sublattices, which facilitate the tuning of relevant strengths of the spin glass and AFM order parameters, thereby providing a new strategy to tailor the EB in future electronic applications.

\section{Experimental details}

Chemical vapor transport method was employed to grow the single crystals of TaFe$_{1.25}$Te$_3$ in a double-zone tube furnace\cite{TaTe1.25Te3,ChenXH}. Firstly, the polycrystalline precursor of TaFe$_{1.25}$Te$_3$ was pre-reacted by melting the stoichiometric amounts of high-purity Ta, Fe, Te elementals at 873 K in a muffle furnace. The resultant ingot with a total mass of 0.5 g was then sealed into a evacuated quartz ampoule together with 5 mg TeCl$_4$ powder as the transport agent. The quartz ampoule (approximately 20 cm in length) was heated in a double-zone furnace for one week, with the hot end at 1053 K and the cold end at 1003 K, i.e., a temperature gradient near 2.5 $^\circ$C cm$^{-1}$. Lastly, pieces of dark gray stripe-like TaFe$_{1.25}$Te$_3$ single crystals with the longest lateral dimension of 1-2 mm were harvested (as shown in Fig. 1 inset).

Single-crystal x-ray diffraction measurements were performed at room temperature using the PANalytical x-ray diffractometer with a monochromatic Cu $K$$\alpha$ radiation. The actual chemical composition of the as-grown single crystals was determined using a field-emission scanning electron microscope equipped with the energy-dispersive X-ray spectroscopy (EDS). The dc magnetic susceptibility was measured using the Magnetic Property Measurement System (MPMS) from Quantum Design. The ac susceptibility was measured with the alternating magnetic field of 10 Oe for frequencies up to 500 Hz.

\section{Results}

In light of the literature\cite{TaTe1.25Te3,ChenXH}, TaFe$_{1.2}$Te$_3$ (only 20$\%$ of the interstitials are occupied) crystallizes in the monoclinic P2$_1$/m space group with the lattice parameters $a$=7.4360${\texttt{\AA}}$, $b$=3.6380${\texttt{\AA}}$, $c$=10.0080${\texttt{\AA}}$ and $\beta$=109.1700$^\circ$. Albeit with the excess Fe atoms in our crystals TaFe$_{1.25}$Te$_3$, i.e., 25$\%$ of the interstitials are filled, the structure is still the same as the Fe-deficient TaFe$_{1.2}$Te$_3$ in terms of its XRD. In its structure, the Fe1 atoms tetrahedrally coordinated by four Te atoms form the one-dimensional zigzag rung propagating along the $b$-axis (Fig. 1a). Note that there is another one-dimensional Ta chain extended along the $b$-axis, coordinated by six Te atoms in octahedral configurations, sharing one corner with the neighboring FeTe$_4$ tetrahedra (Fig. 1a). Additionally, the excess amount of Fe (which we refer to as Fe2, $x$=0.25) partially fill the interstitial sites that are located at the center of square pyramid formed by five Te atoms (Fig. 1b). As illustrated in Fig. 1c, the top/bottom facets of the as-grown crystals are perpendicular to the [1 0 -1] direction. For convenience, we specify two new axes $a'$ and $c'$ that point along the [1 0 1] and [1 0 -1] respectively, such that the $a'$-axis is anchored in the plane and perpendicular to the one-dimensional $b$-axis chains, and the $c'$-axis is normal to the sample top facet. The single crystal XRD patterns (Fig. 1d) only show the reflections from ($\ell$ 0 -$\ell$) peaks and the full width at half maximum (FWHM) of these peaks is less than 0.1$^\circ$, indicating the high degree of crystality of the samples studied.

\begin{figure*}
\includegraphics[width=14cm]{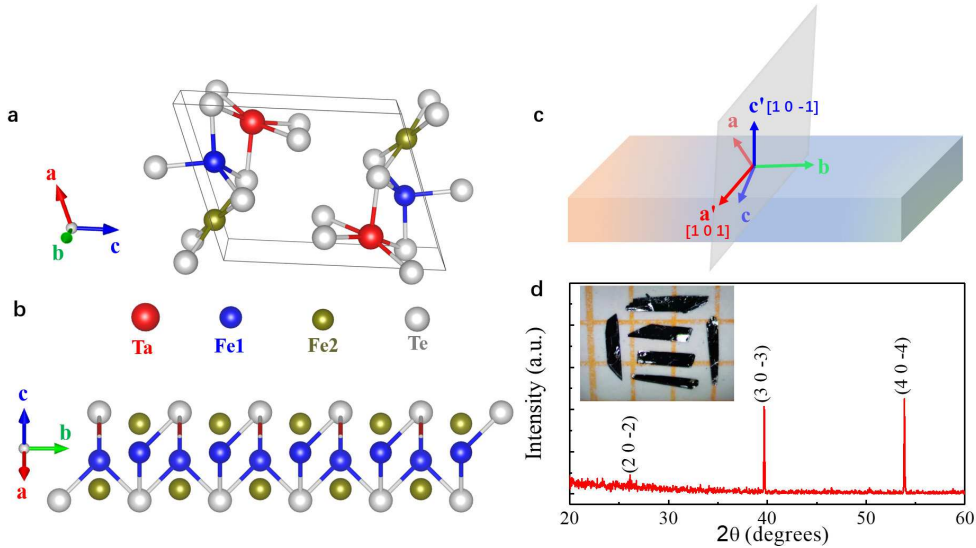}
\caption{\label{fig1} (a) Schematic crystal structure of the Fe2-site fully filled TaFe$_2$Te$_3$. The box indicates one unit cell. (b) The perspective view of TaFe$_2$Te$_3$ structure along the $b$-axis. The Ta atoms are omitted for clarity reason. In TaFe$_{1+x}$Te$_3$, Fe2 sites are only partially occupied. (c) The definition of the axes used in this paper. $c'$ is along [1 0 -1] and perpendicular to the sample top surface and $a'$ (along [1 0 1] direction) lies in the sample's flat plane, normal to the $b$-axis zigzag chain direction. (d) X-ray diffraction pattern from the basal plane of the cleaved crystal, showing only ($l$ 0 -$l$) reflections. The inset shows the optical image of single crystals used in this study.}
\end{figure*}

\begin{figure*}
\includegraphics[width=16cm]{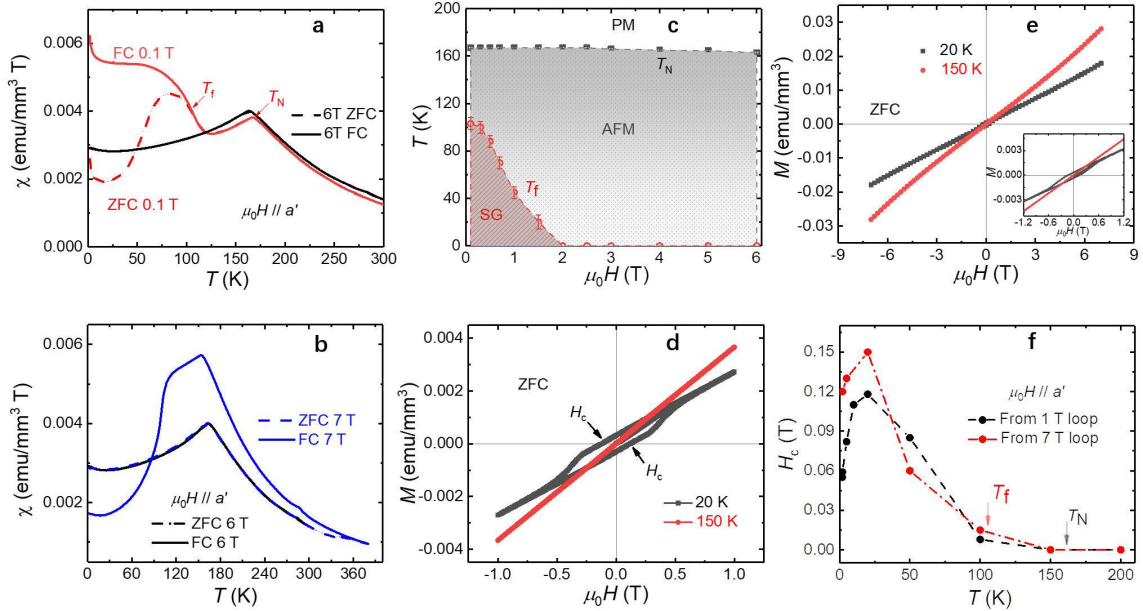}
\caption{\label{fig1}(a) The susceptibility under both ZFC and FC protocols with the field (0.1 T and 6 T) applied along the $a'$-axis. $T_N$ and $T_f$ mark the N$\acute{e}$el temperature and spin glass freezing temperature, respectively. (b) The same as in (a) under a field of 6 T and 7 T. Remarkably, ZFC and FC curves under 7 T start to separate below $\sim$350 K. The reason for this reentrant-like spin glass at such a high field is not clear to us and warrants more investigations in future work. (c) The extracted phase diagram with $H\parallel a'$. PM, AFM and SG stand for paramagnetic, antiferromagnetic and spin glass states, respectively. (d) The magnetization loop (0 T$\rightarrow$1 T$\rightarrow$-1 T$\rightarrow$1 T after ZFC at 20 K (below $T_f$) and 150 K (above $T_f$). $H_c$, the coercive field. (e) The same as in (d) with bigger cycling field (7 T). The inset zooms in on the low field region. (f) $H_c$ from both 1 T and 7 T loops, as a function of temperature, under this field configuration ($H\parallel a'$).}
\end{figure*}

The comprehensive magnetization measurements with field along three different axes $a'$, $b$ and $c'$ reveal the coexisting antiferromagnetic and spin glass orders and their interplay, as we will elaborate in the following. In a low field along the $a'$-axis (0.1 T), the temperature dependence of dc susceptibility $\chi$ (=$M/H$) shows a pronounced kink at $T_N$=167 K, indicating the antiferromagnetic order (Fig. 2a). As the temperature decreases, the ZFC and FC branches begin to separate below $T_f$$\sim$103 K, suggestive of a spin-glassy phase at low temperatures. This bears certain similarities to the Fe-rich or Mn-rich Fe$_x$Mn$_{1-x}$TiO$_3$ where long-range AFM and the reentrant spin glass coexist\cite{Ito88,Ito89,Ito90,Ito93}. With increasing field, $T_f$ is progressively suppressed while $T_N$ is barely changed with the applied field (see Supplemental Fig. 3 in Ref. \cite{SM}). In a field of 2 T, $T_f$ is quenched to zero, evidenced by the coincidence of ZFC and FC curves in the whole temperature range. This reversibility of ZFC and FC curves persists up to 6 T, above which they become separated again. As seen, in a 7 T field, ZFC and FC curves begin to bifurcate at a much higher temperature, $\sim$350 K although the ZFC branch overlaps with the 6 T data below 300 K (Fig. 2b). This irreversibility between ZFC and FC under 7 T at this high temperature is exceptional and seems to suggest the reentrant spin glass at this high field. This re-entrant spin glass behavior in such a high field is surprising and only appears when $H\parallel a'$ and $H\parallel c'$ \cite{SM}, which merits more investigations in the future. The corresponding phase diagram with this field configuration is summarized in Fig. 2c (see also Ref. \cite{SM} for the other two directions).

\begin{figure*}
\includegraphics[width=16cm]{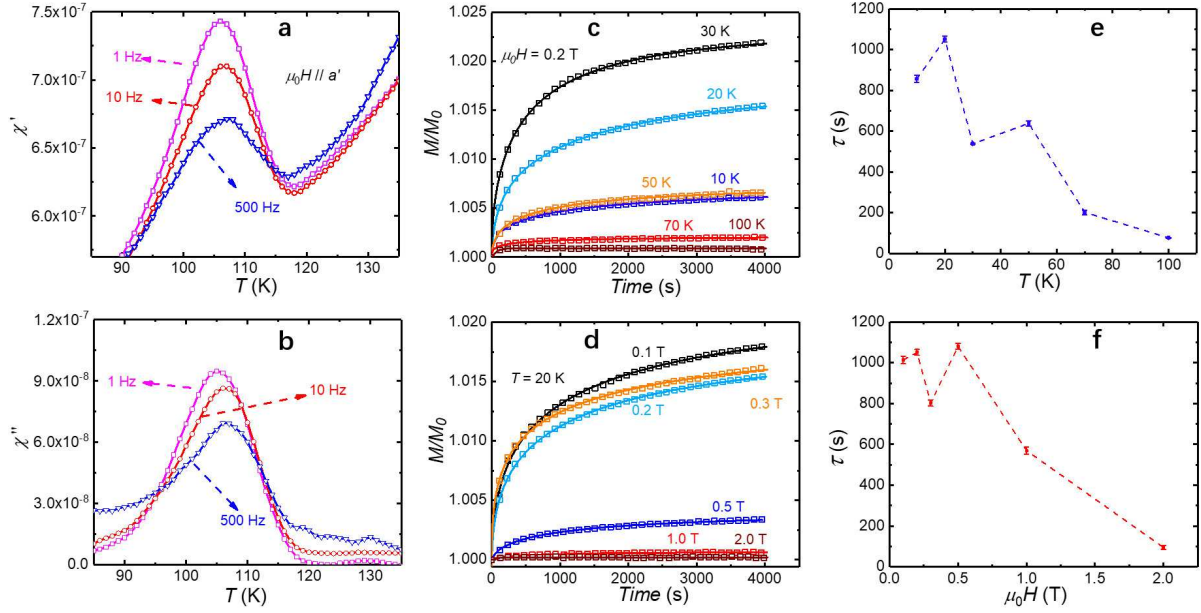}
\caption{\label{fig1} The real (a) and the imaginary (b) parts of the ac susceptibility measured under the ac field (along the $a'$-axis) of 10 Oe. (c) The time dependence of magnetization under a field of 0.2 T. The sample was zero field cooled from above $T_N$ to each temperature and subsequently the field was applied. (d) The time dependence of magnetization at 20 K under various fields. The sample was zero field cooled from above $T_N$ to 20 K in each run. (e) and (f) The relaxation time extracted from panels (c) and (d), respectively.}
\end{figure*}

The spin glass characteristics below $T_f$ were further confirmed by the ac susceptibility and the time dependence of magnetization measurements on TaFe$_{1.25}$Te$_3$ (Fig. 3). The real and imaginary parts of ac susceptibility show a pronounced peak structure at $\sim$$T_f$, which is slightly shifted with the varying frequency, a hallmark of the spin glass state. The time dependence of magnetization below $T_f$ was measured by cooling the sample in zero field from above $T_N$ in each run and the designated field was applied after cooling to each specified temperature. As shown in Fig. 3c and 3d, the relaxation behaviors were studied by measuring $M$ as a function of time, at fixed temperatures or fields. The relaxation of the magnetization can be captured by the Kohlrausch-Williams-Watts decay law\cite{KWW-PRL}:

\begin{equation}
\begin{aligned}
M(t)/M_0=a+b\texttt{exp}[-(t/\tau)^\beta]
\end{aligned}
\end{equation}

\noindent where $a$ and $b$ are constants, $\tau$ is the characteristic relaxation time constant, $\beta$ is the shape parameter and $t$ is the time. The extracted time constants $\tau$ are given in panels Fig. 3e and 3f. Both the ac susceptibility and time dependence bolster the arguments for the spin glass formation below $T_f$.

The hysteresis loop at low temperatures further reveal the spin dynamics associated with this spin glass. As shown in Fig. 2d, the hysteresis loop (0$\rightarrow$1$\rightarrow$-1$\rightarrow$1 T) is manifest at 20 K (below $T_f$) but is absent at 150 K (above $T_f$). This is the same when the loop is conducted in a larger loop (0$\rightarrow$7$\rightarrow$-7$\rightarrow$7 T, Fig. 2e). It is noteworthy that the loop is symmetric when the sample is zero-field-cooled, i.e., there is no ZFC EB in this material, in sharp contrast to the loop when the sample is field-cooled, as we will discuss later. The coercive field $H_c$, defined as the field at which the magnetization goes to zero, is plotted in Fig. 2f. Only below $T_f$, as seen, does $H_c$ start to grow and reach a peak at $\sim$20 K. As noted, the 1 T loop and 7 T loop show the same dependence of $H_c$ on temperature with slightly different values.

The same treatments with field along the other two directions provide more insights into the spin configuration and its dynamics (Supplemental Fig. 1 and 2 in Ref. \cite{SM}). With $H$$\parallel$$b$, the antiferromagnetic transition only manifest itself as a small kink (Supplemental Fig. 1 in Ref. \cite{SM}), compared to the large drop in the other two configurations, although the susceptibility shows only weak anisotropy for the three directions (as a reference, $\chi_{a'}$:$\chi_b$:$\chi_{c'}$=1:1.55:1.5 at $T_N$). Such anisotropy is commonly observed in layered materials with divergent susceptibilities and is attributed to anisotropic Lande $g$ factors \cite{Terasaki}. From symmetry arguments, this contrasting behavior at $T_N$ along three field directions implies that the spins are predominantly aligned in the $ac$ plane, with only a small component along the chain\cite{Hussey}. With $H$$\parallel$$b$, the irreversibility is also observed between ZFC and FC at low fields. Unlike the other two directions, the susceptibility at 7 T shows no re-entrant spin glass behavior for $H$$\parallel$$b$ (Supplemental Fig. 1 in Ref. \cite{SM}). The origin of this difference is, however, not clear. Remarkably, the hysteresis loop is not seen in $H$$\parallel$$b$ whereas it is present in $H$$\parallel$$c'$ and $H$$\parallel$$a'$, in concert with our identification that spins are predominantly aligned in the $ac$ plane, with only a small in-chain component. This picture was also corroborated by the elastic neutron scattering measurements where it was suggested that the intra-ladder Fe1 spins are ferromagnetically ordered but the inter-ladder coupling between these ladders are antiferromagnetic essentially, with the interstitial Fe2 spins aligned parallel to their neighboring Fe1 spins\cite{Ke}. $H_c$ for field along the $c'$-axis at different temperatures is depicted in Supplemental Fig. 2 in Ref. \cite{SM}, which is present only below the spin glass ordering $T_f$. Consequently, it is conceivable to attribute the origin of this hysteresis to its spin glass phase.

\begin{figure*}
\includegraphics[width=16cm]{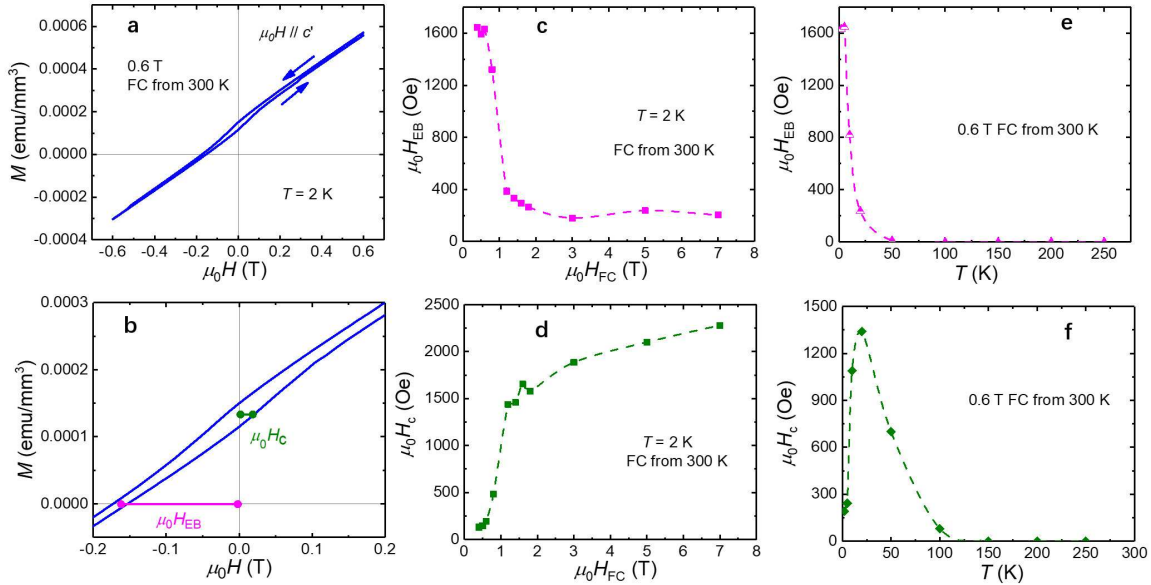}
\caption{\label{fig1}(a) Shifted magnetic hysteresis loop at 2 K, after the sample was field cooled in 0.6 T from room temperature. The field was applied along the $c'$ axis. (b) The zoomed-in view of (a). The exchange bias $H_{EB}$ is defined as the average of $x$ intercepts. The coercive field $H_c$ is a measure of half-width of the hysteresis loop at the average of the $y$ intercepts. (c) and (d) The extracted $H_{EB}$ and $H_c$ at 2 K as a function of cooling field $H_{FC}$ when the sample was cooled from room temperature. (e) and (f) The temperature dependence of $H_{EB}$ and $H_c$ respectively, as the sample was cooled in 0.6 T field from 300 K to that temperature. }
\end{figure*}

The exchange coupling between antiferromagnetic and spin glass orders is evidenced from the hysteresis loop that is strongly shifted when the sample is cooled in field from above $T_N$, the so-called exchange bias effect. We exemplify the hysteresis loop when the sample is field-cooled in 0.6 T ($H\parallel c'$) from 300 K and field-cycled at 2 K (Fig. 4a). The loop is obviously shifted along the lateral direction, as magnified in Fig. 4b. Here although we cannot completely dismiss the minor loop effect as the origin of this hysteresis shift at 0.6 T because the field is not strong enough to saturate the magnetization\cite{Gesheva-comment}, the robust bias after several training loops seems at odds with the minor loops. Moreover, if the displacement originates from a minor loop effect, the ZFC curve should be shifted as well\cite{Felser-IrMnGa}, which is not the case we observed (see ZFC hysteresis loop at 2 K in Supplemental Fig. 2d with $H\parallel c'$ where no displacement is found \cite{SM}). Additionally, at higher fields when the hysteresis loops become closed, the displacement of the loop is still present (For example, when $\mu_0$$H_{FC}$= 5 T, $H_{EB}$ = 240 Oe). All these facts seem inconsistent with the minor loop effect as the origin of this displacement. Next, we define the exchange bias field $H_{EB}$ as the average of the field at which the magnetization curve intercepts the $x$-axis (field axis), and the coercive field $H_c$ as the half-width of the hysteresis loop at the average of the $y$-intercepts (the $M$-axis) (see Fig. 4b). Both $H_{EB}$ and $H_c$ are not only dependent on the cooling field $H_{FC}$ (under which the sample is cooled), but also strongly varying with the temperature at which the hysteresis loop is cycled. It is notable that $H_{EB}$ peaks at low field and decreases quickly with increasing field and finally plateaus above 3 T (Fig. 4c). At 5 T (where the spin glass order is very weak in this field direction, if at all), $H_{EB}$ retains a magnitude of 240 Oe. The field dependence of $H_c$ show the opposite trend to $H_{EB}$; it increases with increasing cooling field and shows no sign of saturation at 7 T (Fig. 4d). As seen from Fig. 4e, $H_{EB}$ is quickly quenched at temperature above 50 K. This seems to suggest that although the spin glass order sets on at $\sim$100 K, it is only well established at temperature of 50 K below which the exchange coupling of antiferromagnetic and spin glass orders start to proliferate. The temperature dependence of $H_c$, on the other hand, shows similar trend to that in the ZFC mode (Fig. 2f); it increases fast with increasing temperature, peaks at 20 K and drops very quickly, and finally disappears above $T_f$ (Fig. 4f).

\section{Discussion and Conclusion}

The strong coupling between antiferromagnetic order and spin glass evidenced from our experiments may have profound implications for its applications in spintronic device. Hitherto, a handful of antiferromagnets have been reported whose spin texture can be switched by electrical current via the spin-orbit torque\cite{Wadley-Science,Bodnar-NC,Analytis-NM,Orenstein-NM,Analytis-SA}. In these materials, due to the broken inversion symmetry and spin orbit coupling, the applied electrical current are spin polarized and will exert a spin-orbit torque on the existing antiferromagnetic spins when the spin polarized current flows through, which therefore transfer the angular momentum into the spin system and switch the magnetic domains of the existing AFM\cite{JMMM,Gomonay-PRB,Jungwirth-PRL}. This is the mechanism of electrical switching in antiferromagnets. However, the switching efficiency in most antiferromagnets is low. The coexistence of spin glass and antiferromagnetism, however, opens the new channel for transferring angular momentum to the system, thereby enhancing the efficacy of switching. That, in turn, leverage the local stiffness of the spin glass. This cooperative interplay between antiferromagnetic and spin glass orders has recently been seen in the disordered Fe$_{1/3+\delta}$NbS$_2$\cite{Analytis-SA}. In analogy to Fe$_{1/3+\delta}$NbS$_2$, the collective spin dynamics below $T_f$ in TaFe$_{1.25}$Te$_3$ can also develop a new channel for transferring spin and impart the angular momentum to AFM, facilitating the rapid switching of antiferromagnetism by electrical current \cite{Analytis-SA}.

Compared to Fe$_{1/3+\delta}$NbS$_2$, TaFe$_{1+x}$Te$_3$ system has its own advantages in the electrical switching study. First, in Fe$_{1/3+\delta}$NbS$_2$, AFM and spin glass both derive from the same triangular lattice of iron sandwiched by NbS$_2$ layers. Nevertheless, in TaFe$_{1+x}$Te$_3$, AFM and spin glass originate from two different sublattices, the former from the zigzag Fe1 ladder and the latter from the interstitial Fe2, which make it possible to tailor the strength of spin glass without affecting the AFM order parameter. Second, the Fe concentration in TaFe$_{1+x}$Te$_3$ can be tuned in a large range, at least from $x$=0.2 to $x$=0.6, offering more flexibility in this tuning. Third, TaFe$_{1+x}$Te$_3$ possesses Q1D spin chains, which offers an ideal platform to study the interplay among magnetism, electrical switching and dimensionality in a bulk material system.

In summary, we have studied the coexistence of AFM and spin glass orders in the Q1D iron telluride TaFe$_{1.25}$Te$_3$ and the exchange bias effect due to their coupling, which was not reported in its homologues with fewer iron intercalations ($x$=0.17 \cite{Ke} and 0.21 \cite{ChenXH}). The concerted efforts from the collective dynamics of anisotropic antiferromagnetism and spin glass revealed in this study not only leads to the exchange bias orders of magnitude larger than the typical exchange-bias systems, more importantly it may also provide a new platform to manipulate or switch the magnetism in a controllable way. Looking forward, it would be interesting to tune the possible superconductivity \cite{EuFeAsP-PRL,BaFeCoAs-PRL} or other unconventional electronic states in this spin-glass iron-based chalcogenide family by hydrostatic pressure or electrostatic gating which is currently envisaged.

\begin{acknowledgments}

The authors would like to thank C. M. J. Andrew, Xin Lu for stimulating discussions. This work is sponsored by the National Natural Science Foundation of China (Grant No. 11974061, No. 12004337, No. 51861135104), the National Key Research and Development Program of China (Grant No. 2016YFA0401704), and by Zhejiang Provincial Natural Science Foundation of China (No. Q21A040024). X. Ke acknowledges the financial support from the start-ups at Michigan State University.

Y. Liu, J. J. Bao and C. Q. Xu contributed equally to this work.

\end{acknowledgments}

\appendix


\begin{thebibliography}{199}

\bibitem{SBlundell}  S. Blundell, Magnetism in Condensed Matter, \emph{Oxford Express}.

\bibitem{Felser-IrMnGa} J. Kroder, K. Manna, D. Kriegner, A. S. Sukhanov, E. Liu, H. Borrmann, A. Hoser, J. Gooth, W. Schnelle, D. S. Inosov, G. H. Fecher, and C. Felser, Spin glass behavior in the disordered half-Heusler compound IrMnGa, \emph{Phys. Rev. B.} \textbf{99}, 174410 (2019).

\bibitem{Chatterjee} S. Chatterjee, S. Giri, S. K. De, and S. Majumdar, Reentrant-spin-glass state in Ni$_2$Mn$_{1.36}$Sn$_{0.64}$ shape-memory alloy, \emph{Phys. Rev. B.} \textbf{79}, 092410 (2009).

\bibitem{APL-WuGH} L. Ma, W. H. Wang, J. B. Lu, J. Q. Li, C. M. Zhen, D. L. Hou, and G. H. Wu, Coexistence of reentrant-spin-glass and ferromagnetic martensitic phases in the Mn$_2$Ni$_{1.6}$Sn$_{0.4}$ Heusler alloy, \emph{Appl. Phys. Lett.} \textbf{99}, 182507 (2011).

\bibitem{Analytis-NP} E. Maniv, R. A. Murphy, S. C. Haley, S. Doyle, C. John, A. Maniv, S. K. Ramakrishna, Y. Tang, P. Ercius, R. Ramesh, A. P. Reyes, J. R. Long and J. G. Analytis, Exchange bias due to coupling between coexisting antiferromagnetic and spin-glass orders, \emph{Nat. Phys.} \textbf{17}, 525 (2021).

\bibitem{Analytis-NC} E. Lachman, R. A. Murphy, N. Maksimovic, R. Kealhofer, S. Haley, R. D. McDonald, J. R. Long and J. G. Analytis, Exchange biased anomalous Hall effect driven by frustration in a magnetic kagome lattice, \emph{Nat. Commun.} \textbf{11}, 560 (2020).

\bibitem{BMWang-PRL} B. M. Wang, Y. Liu, P. Ren, B. Xia, K. B. Ruan, J. B. Yi, J. Ding, X. G. Li, and L. Wang, Large Exchange Bias after Zero-Field Cooling from an Unmagnetized State, \emph{Phys. Rev. Lett.} \textbf{106}, 077203 (2011).

\bibitem{Felser-PRL} A. K. Nayak, M. Nicklas, S. Chadov, C. Shekhar, Y. Skourski, J. Winterlik, and C. Felser, Large Zero-Field Cooled Exchange-Bias in Bulk Mn$_2$PtGa, \emph{Phys. Rev. Lett.} \textbf{110}, 127204 (2013).

\bibitem{Felser-NM} A. K. Nayak, M. Nicklas, S. Chadov, P. Khuntia, C. Shekhar, A. Kalache, M. Baenitz, Y. Skourski, V. K. Guduru, A. Puri, U. Zeitler, J. M. D. Coey and C. Felser, Design of compensated ferrimagnetic Heusler alloys for giant tunable exchange bias, \emph{Nat. Mater.} \textbf{14}, 679 (2015).

\bibitem{Murthy-APL} J. Krishna Murthy and A. Venimadhav, Giant zero field cooled spontaneous exchange bias effect in phase separated La$_{1.5}$Sr$_{0.5}$CoMnO$_6$, \emph{Appl. Phys. Lett.} \textbf{103}, 252410 (2013).

\bibitem{Boldrin-APL} M. Boldrin, A. G. Silva, L. T. Coutrim, J. R. Jesus, C. Macchiutti, E. M. Bittar, and L. Bufai\c{c}al, Tuning the spontaneous exchange bias
effect with Ba to Sr partial substitution in La$_{1.5}$(Sr$_{0.5-x}$Ba$_x$)CoMnO$_6$, \emph{Appl. Phys. Lett.} \textbf{117}, 212402 (2020).

\bibitem{Meiklejohn} W. H. Meiklejohn and C. P. Bean,  New magnetic anisotropy, \emph{Phys. Rev.} \textbf{105}, 904 (1957).

\bibitem{Cain} W. C. Cain and M. H. Kryder, Investigation of the exchange mechanism in NiFe-TbCo bilayers. \emph{J. Appl. Phys.} \textbf{67}, 5722 (1990).

\bibitem{Ali-NM} Ali, M. et al. Exchange bias using a spin glass, \emph{Nat. Mater.} \textbf{6}, (2007).


\bibitem{Ke} X. Ke, B. Qian, H. Cao, J. Hu, G. C. Wang, and Z. Q. Mao, Magnetic structure of quasi-one-dimensional antiferromagnetic TaFe$_{1+y}$Te$_3$, \emph{Phys. Rev. B.} \textbf{85}, 214404 (2012).


\bibitem{TaTe1.25Te3} M. E. Badding, J. Li, F. J. Disalvo, W. Zhou and P. P. Edwards, Characterization of TaFe$_{1.25}$Te$_3$, a New Layered Telluride with an
Unusual Metal Network Structure, \emph{J. Solid State Chem.} \textbf{100}, 313 (1992).


\bibitem{ChenXH} R. H. Liu, M. Zhang, P. Cheng, Y. J. Yan, Z. J. Xiang, J. J. Ying, X. F. Wang, A. F. Wang, G. J. Ye, X. G. Luo, and X. H. Chen, Spin-density-wave transition of Fe1 zigzag chains and metamagnetic transition of Fe2 in TaFe$_{1+y}$Te$_3$, \emph{Phys. Rev. B.} \textbf{84}, 184432 (2011).

\bibitem{Ito88} H. Ragura, A. Ito, H. Wakabayashi, T. Goto, Field-dependent phenomena in reentrant-spin-glass Fe$_x$Mn$_{1-x}$TiO$_3$ with $x$=0.60, 0.65 and 0.75, \emph{J. Phys. Soc. Jpn} \textbf{57}, 2636 (1988).

\bibitem{Ito89} H. Yoshizawa, S. Mitsuda, H. Aruga, A. Ito, Reentrant spin glass transition and a mixed phase in an Ising system Fe$_x$Mn$_{1-x}$TiO$_3$, \emph{J. Phys. Soc. Jpn} \textbf{58}, 1416 (1989).

\bibitem{Ito90} A. Ito, S. Morimoto and H. Aruga, Studies On spin-glass freezing and antiferromagnetic long-range order in an Ising spin-glass system Fe$_x$Mn$_{1-x}$TiO$_3$, \emph{Hyperfine interactions} \textbf{54}, 567 (1990).

\bibitem{Ito93} H. Aruga Katori and A. Ito, Magnetic property and phase diagram of a frustrated system with competing exchange interactions, Fe$_x$Mn$_{1-x}$TiO$_3$, \emph{J. Phys. Soc. Jpn} \textbf{62}, 4488 (1993).

\bibitem{SM} See Supplemental Material for the phase diagrams along $H\parallel b$ and $H\parallel c'$, and the magnetization curves under various fields along all three directions ($H\parallel a'$, $H\parallel b$ and $H\parallel c'$).

\bibitem{KWW-PRL} J. Hessinger and K. Knorr, Field-Cooling Experiments on the Quadrupolar-Glass State of (KBr)$_{0.47}$(KCN)$_{0.53}$, \emph{Phys. Rev. Lett.} \textbf{65}, 2674 (1990).


\bibitem{Terasaki} I. Terasaki, M. Hase, A. Maeda, K. Uchinokura, T. Kimura, K. Kishio, I. Tanaka, and H. Kojima, Doping effects on the anisotropic magnetic susceptibility in single-crystal La$_{2-x}$Sr$_x$CuO$_4$, \emph{Physica C} \textbf{193}, 365 (1992).

\bibitem{Hussey} X. Xu, A. Carrington, A. I. Coldea, A. Enayati-Rad, A. Narduzzo, S. Horii, and N. E. Hussey, Dimensionality-driven spin-flop transition in quasi-one-dimensional PrBa$_2$Cu$_4$O$_8$, \emph{Phys. Rev. B} \textbf{81}, 224435 (2010).


\bibitem{Gesheva-comment} J. Gesheva, Comment on Cluster glass induced exchange biaslike effect in the perovskite cobaltites, \emph{Appl. Phys. Lett.} \textbf{93}, 176101 (2008).

\bibitem{Wadley-Science} P. Wadley, B. Howells, J. $\check{Z}$elezny, C. Andrews, V. Hills, R. P. Campion, V. Nov$\acute{a}$k, K. Olejn$\acute{i}$k, F. Maccherozzi, S. S. Dhesi, S. Y. Martin, T.Wagner, J. Wunderlich, F. Freimuth, Y. Mokrousov, J. Kune$\breve{s}$, J. S. Chauhan, M. J. Grzybowski, A. W. Rushforth, K.W. Edmonds, B. L. Gallagher, T. Jungwirth, Electrical switching of an antiferromagnet, \emph{Science} \textbf{351}, 587 (2016).

\bibitem{Bodnar-NC} S.Yu. Bodnar, L. $\check{S}$mejkal, I. Turek, T. Jungwirth, O. Gomonay, J. Sinova, A.A. Sapozhnik, H.-J. Elmers, M. Kl$\ddot{a}$ui and M. Jourdan, Writing and reading antiferromagnetic Mn$_2$Au by N$\acute{e}$el spin-orbit torques and large anisotropic magnetoresistance, \emph{Nat. Commun.} \textbf{9}, 348 (2018).

\bibitem{Analytis-NM} N. L. Nair, E. Maniv, C. John, S. Doyle, J. Orenstein and J. G. Analytis, Electrical switching in a magnetically intercalated transition metal dichalcogenide, \emph{Nat. Mater.} \textbf{19}, 153 (2020).

\bibitem{Orenstein-NM}  A. Little, C. Lee, C. John, S. Doyle, E. Maniv, N. L. Nair, W. Chen, D. Rees, J$\ddot{o}$rn W. F. Venderbos, R. M. Fernandes, J. G. Analytis and J. Orenstein, Three-state nematicity in the triangular lattice antiferromagnet Fe$_{1/3}$NbS$_2$, \emph{Nat. Mater.} \textbf{19}, 1062 (2020).

\bibitem{Analytis-SA}  E. Maniv, N. L. Nair, S. C. Haley, S. Doyle, C. John, S. Cabrini, A. Maniv, S. K. Ramakrishna, Y. Tang, P. Ercius, R. Ramesh, Y. Tserkovnyak, A. P. Reyes, J. G. Analytis, Antiferromagnetic switching driven by the collective dynamics of a coexisting spin glass, \emph{Sci. Adv.} \textbf{7}, eabd8452 (2021).


\bibitem{JMMM} D. C. Ralpha and M. D. Stiles, Spin transfer torques, \emph{J. Magn. Magn. Mater.} \textbf{320}, 1190 (2008).

\bibitem{Gomonay-PRB} H. V. Gomonay, V. M. Loktev, Spin transfer and current-induced switching in antiferromagnets, \emph{Phys. Rev. B.} \textbf{81}, 144427 (2010).

\bibitem{Jungwirth-PRL} J. $\check{Z}$elezn$\acute{y}$, H. Gao, K. V$\acute{y}$born$\acute{y}$, J. Zemen, J. Ma$\check{s}$ek, Aur$\acute{e}$lien Manchon, J. Wunderlich, J. Sinova, and T. Jungwirth, Relativistic N$\acute{e}$el-Order Fields Induced by Electrical Current in Antiferromagnets, \emph{Phys. Rev. Lett.} \textbf{81}, 144427 (2010).

\bibitem{EuFeAsP-PRL} S. Zapf, H. S. Jeevan, T. Ivek, F. Pfister, F. Klingert, S. Jiang, D. Wu, P. Gegenwart, R. K. Kremer, and M. Dressel, EuFe$_2$(As$_{1-x}$P$_x$)$_2$: Reentrant Spin Glass and Superconductivity, \emph{Phys. Rev. Lett.} \textbf{110}, 237002 (2013).

\bibitem{BaFeCoAs-PRL} A. P. Dioguardi, J. Crocker, A. C. Shockley, C. H. Lin, K. R. Shirer, D. M. Nisson, M. M. Lawson, N. apRoberts-Warren, P. C. Canfield, S. L. Bud'ko, S. Ran, and N. J. Curro, Coexistence of Cluster Spin Glass and Superconductivity in Ba(Fe$_{1-x}$Co$_x$)$_2$As$_2$ for 0.060$\leq$$x$$\leq$0.071, \emph{Phys. Rev. Lett.} \textbf{111}, 207201 (2013).







\end{thebibliography}
\end{document}